# Redesigning the Urban Design Studio: Two Learning Experiments

**Burak Pak**
burak.pak@kuleuven.be

**Johan Verbeke**
Johan.Verbeke@kuleuven.be

Faculty of Architecture
KU Leuven University, Brussels

**Abstract**

*The main aim of this paper is to discuss how the combination of Web 2.0, social media and geographic technologies can provide opportunities for learning and new forms of participation in an urban design studio. This discussion is mainly based on our recent findings from two experimental urban design studio setups as well as former research and literature studies. In brief, the web platform enabled us to extend the learning that took place in the design studio beyond the studio hours, to represent the design information in novel ways and allocate multiple communication forms. We found that the students' activity in the introduced web platform was related to their progress up to a certain extent. Moreover, the students perceived the platform as a convenient medium and addressed it as a valuable resource for learning. This study should be conceived as a continuation of a series of our "Design Studio 2.0" experiments which involve the exploitation of opportunities provided by novel socio-geographic information and communication technologies for the improvement of the design learning processes.*

**Keywords**
*Design Studio 2.0, e-learning, peer learning, collective mapping, affordances*

## Introduction and Background

During the last decade, the convergence of Web 2.0, social media and geographic technologies resulted in the development of innovative knowledge production tools and strategies which facilitate collaborative and location-aware learning (Lund, 2013). Through the use of these technologies, it is possible to enhance our powers of observation, create richer and authentic learning experiences in which the learners collaborate in creating new knowledge and extend their own understandings (Lloyd, 2010). They can potentially enable new constructivist learning modes: particularly in socio-spatially situated and media-rich learning contexts such as the architectural and urban design studios. When combined with novel learning strategies, these technologies can promote and augment rigorous discussion and informed consensus on actions and design problems (Schnabel & Ham, 2011).

In this context, the main purpose of this paper is to discuss how the combination of Web 2.0, social media and geographic technologies can provide opportunities for learning and new forms of participation in an urban design studio. An important aspect will be the fundamental properties that determine how these could possibly be used, in other words, their perceived affordances (Norman, 1988, p. 9). This discussion will be mainly based on our recent findings from two experimental urban design studio setups as well as former research and literature studies.

Considering that social participation is a process of learning and knowing (Wenger, 1998), we will question the nature of learning and participation that takes place in it, drawing on the learning experiences of the students as well as the traces of their online activity logs. Before starting to reveal the details on our study, we find it useful to briefly share a few relevant domain-specific characteristics and observations that motivated our study, since the target audience of this journal is not limited to architectural or urban designers.

## The Design Studio

The educational setting of the architectural schools depends on highly reflective practices which focus on the "design studio" as their central component (Schön, 1987). The main reason for this is that design knowledge is difficult to externalise and *tacit* in essence (Polanyi, 1966). As a response to this challenge, beginning with the nineteenth century ateliers of the *École des Beaux-Arts,* the design studio-learning environment has emerged and slowly evolved (Scheeler & Smith, 2007).

Today, the design studio is understood as a place where students learn experientially by designing their own projects through periodical critiques and collective reviews. In this sense, the design studio has a significantly different setup compared to the theoretical academic courses and seminars (Schwendimann, 2013). During the critiques, learners "seek to dwell in the moves of" an experienced designer (teacher) and "try to understand it by observation, imitation and picking out the essential features of the action" (Polanyi, 1966, p. 30), in other words they "know-in-action" (Schön, 1987, p. 22).

Through the reflective processes of drawing and modelling, students develop design alternatives and interpret and explore their impacts relying on representations and social/self-reflections. The students are expected to consider their design alternatives together with the existing social and spatial urban environment and build relationships between these while redefining them.

The learning that takes place in the design studio is not only limited to the interactions between the teacher and the learner. Due to the nature and complexity of the design problems, disciplinary and multidisciplinary collaboration and reflection are essential competences that the students need to acquire (Kendall, 2007). Ideally, the design studio should involve social knowledge building through design actions and reflections between the students (peers) in the form of internal and external conversations with the design situation; as well as the socio-spatial context; in other words "reflection-in-action" (Schön, 1987, p. 28).

However, in contemporary educational practices, the potential of peer learning is not always fully exploited; partially because the communication between student and studio teacher is a problematic and complex one (Schön, 1987). Although the authors recognise the diversity of teaching practices around the world, various studies from different countries have reported that the peer learning aspect can be undermined due to an overemphasis on the teacher, rather than on the student (Koch et al., 2002; Newton & Boie, 2011; Webster, 2007).

Another potential threat to this form of constructivist learning in the design studio is a possible overemphasis on the final design outcome. The moments of collective critique channel the students to work long hours towards a presentation in front of a jury of experts. These one-off occasions organised in a limited amount of time divert the focus towards the product rather than the whole learning process (Koch et al., 2002) and create a "skewed" power hierarchy in which students have to justify their work and thoughts in a spatial jury setting that is reinforcing this hierarchical relation. In addition, due to the time consuming nature of the critiques, very little time is left to the students to comment on and criticise each other's works, and they are not motivated enough to be active participants in the discussion (Webster, 2007).

### *Design Studio 2.0*

In the context described above, Web 2.0-based social software and geospatial technologies can be seen as potential platforms to remediate and extend the reflective conversation between the teachers and students in the design studios. They can be designed to enable reflective learning-in-action in the design studio in a novel pedagogical context, in which various communication modes and styles are supported. Inspired by the definition of Web 2.0 (O'Reilly, 2005) we call studio learning environments that involve such practices "Design Studio 2.0."

Design Studio 2.0 differs from the classical design studio described above in terms of available communication modes and styles, learning experiences, studio focus, studio environment, time, information resources and representation of design information (Pak & Verbeke, 2012). This may be summarised as follows:

- Design Studio 2.0 can make way to blended learning which refers to the combination of conventional and online learning activities.
- Compared with the conventional studio, the focus of the Design Studio 2.0 is more oriented towards the students, the design products and the learning processes.
- Architects operate in a virtual world, a constructed representation of the real world of practice (Schön, 1987). The Design Studio 2.0 learning environment extends this world to a shared and globally accessible virtual world creating novel potentials for collaboration.
- Learning in the Design Studio 2.0 can take place outside the school environment and is not limited to the studio hours.
- Design Studio 2.0 supports the design information to be represented in novel ways, including the use of 3D models (4D with the inclusion of time), scanned versions of sketches and drawings, computer drawings and renderings, dynamic maps, geolocated notes, and comments.
- Besides the synchronous communication form, asynchronous and combined communication forms can be supported.
- While the conventional design studio involves face-to-face communication, the Design Studio 2.0 also includes avatar-to-avatar communication.

Combining all the possibilities above, we can conclude that the Design Studio 2.0 can offer numerous opportunities that are not fully or easily available in a conventional design studio setting.

### *Former findings and relevant implementations*

The potentials of Design Studio 2.0 have been partially demonstrated by various practical implementations. Burrow and Burry (2006) reported the effective use of Wikis as an internationally distributed design research network incorporating diverse forms of expertise and focusing on the extension of the Sagrada Familia Church in Barcelona. Chase et al. (2008) introduced the "Wikitecture" concept as a decentralised method of open source co-production and tested the use of a three dimensional Wiki to collaboratively develop a design competition entry.

The Housing@21.Eu (2006) project demonstrated the potentials of Web 2.0-based learning environments as an architectural repository and a design inspiration platform. Building on this experience, the OIKODOMOS Project (Madrazo et al., 2013) developed a blended learning pedagogy incorporating a web-based learning space in which teachers and students of schools of architecture and urban planning collaborated in the design and implementation of learning activities dedicated to the study of housing.

Besides the educational domain, various urban design and planning related organisations have developed experimental participatory urban design applications using Web 2.0-based social software and geospatial technologies. Examples of these kinds of initiatives are the Copenhagen

Municipality's *indrebylokaludval* web application, *aloitekanava* by the city of Turku, *"Bristol Rising"* by the Bristol City, *"civic crowd"* sponsored by the British Design Council, *"Change by Us"* by the cities of New York and Philadelphia, *"Spacehive"* by multiple actors in London, *"Lighter Quicker Cheaper"* in San Antonio City, *"mycitylab"* in Brussels and *"Fix My Street"*, *"Neighborland"*, *"SeeClickFix"*, *"Openplans"* (covering multiple cities) which are used for the collection of the ideas from citizens.

In addition to the above, in 2010, we conducted an eight-week long international urban Design studio 2.0 experiment with 39 students in which a geographic MediaWiki was used for the collaborative and location-based analysis of the project site (Pak & Verbeke, 2012).

In this study, we looked for possible impacts of the introduced platform through web use statistics, feedback sessions and a comprehensive questionnaire. The most prominent finding was a strong correlation between online collaborative edits and student results which weakens after a certain threshold (more than 240 group edits in eight weeks). Although the number of participants was not statistically significant for generalisation, this correlation suggested that the use of the platform may have increased students' progress to a certain extent (or vice versa). Furthermore, making edits more than a threshold may as well affect their progress negatively.

## Research Design and Implementation

In order to explore the potentials, possible contributions and challenges of Web 2.0, social media and geographic technologies in learning and participation in design studios, we have designed and tested two experimental setups (Table 1). This design was majorly based on the analysis of relevant implementations (as part of a post-doctoral research project supported by the Brussels Government) (Pak, 2009), our former eight week experience in a similar design studio organised in 2010 and the student feedback received during this study (Pak & Verbeke, 2012).

Table 1

*The context, number of students, duration, aims and affordances in two design studios*

| Context | Students | Duration | Aims | Affordances for learning |
| --- | --- | --- | --- | --- |
| **Design Studio DS1** *International Urban Design Studio* Date: Spring 2012 Location: Brussels Topic: Developing an urban project Project Site: Kirchberg, *Luxembourg* Theme: Eatscape | (*n*= 34) Masters | 14 weeks | Augment the urban design learning by remediating and extending the reflective conversation in the design studio; collectively construct a shared memory on a *remote* urban space | Asynchronous Sharing of Design Information Collective Mapping Commenting on Peers Sharing Design Information External Expert Comments Weekly online challenges and questions (asynchronous) Weekly face to face discussion and two collective critiques |
| **Design Studio DS2** *International Urban Design Studio* Date: Spring 2013 Location: Brussels Topic: Developing urban | (*n*=27) Masters | 14 weeks | Augment the urban design learning by remediating and extending the reflective conversation in the design studio; collectively | Asynchronous Sharing of Design Information Collective Mapping Commenting on Peers' posts Sharing Design Information Rating Peers' projects External Expert Comments Survey and Questionnaire Internal messaging |

| projects<br>Project Site:<br>Administrative city, *Brussels*<br>Theme: Health City | construct a shared memory on a *local* urban space | Location-aware mobile reach<br>Weekly Online challenges and questions (asynchronous)<br>Weekly face-to-face discussion and two collective critiques |

Both of the design studios were organised at (Pak & Verbeke, 2012) and the first author actively participated in the teaching process. Both studios had similar aims and combined online and offline face-to-face learning activities in line with our Design Studio 2.0 description introduced in the previous section.

However, the two studios should be treated as unique cases. It is necessary to recognise that the knowledge and feedback acquired from the former (DS1) were transferred to the latter (DS2). This is why the DS2 includes a wider set of affordances that were also accompanied by technological and usability improvements.

In addition, while the scales of the student projects were similar, the themes were quite different. In DS1 the students used "Eatscape" concept as a designerly lens whereas in DS2 they explored the "Health City" theme. Therefore, while comparing the results, it is necessary to keep in mind the differences between two cases.

### *Rethinking the Design Studio as a Multi-layered process*

In order to organise two "Design Studio 2.0" experiments, it was necessary to reconsider the relations between the courses in the curriculum. Figure 1 shows how the structure of the course has changed.

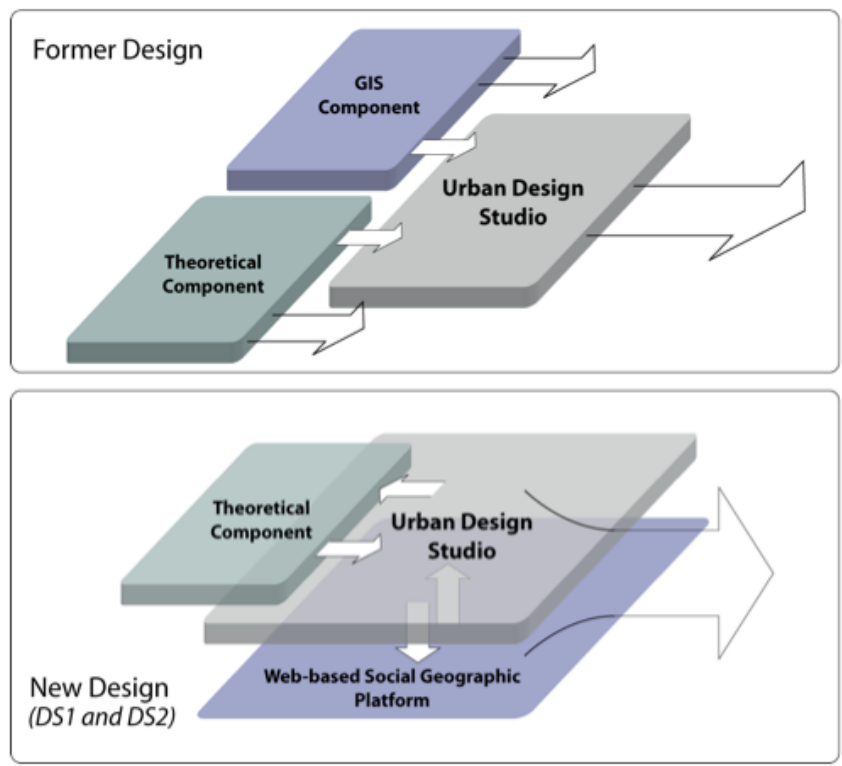

*Figure 1.* The conventional design studio setup in the curriculum (above) and the experimental setup used in DS1 and DS2 (below).

In former years, as shown in the top part of Figure 1, the theoretical and technical courses were designed as individual elements feeding the design studio in a single direction. More recently, as a result of a recent redesign of the curriculum focusing on practice-based learning, the urban design theory component was partially integrated into the design studio. In this setup, the theory teachers take part in the studio classes and motivate the students to reflect what they have learned in theory courses onto their design practices (and vice versa). Formerly, the GIS (Geographic Information System) component was thought as a specialised body of knowledge disconnected from the design studio. Therefore, very few students were able to use this knowledge in their design processes. Informed by this issue, we have reconfigured the urban design studio as a multi-layered process similar to an urban design studio (as shown in the lower part of Figure 1).

In the newer structure, the Web 2.0-based social geographic platform plays an augmenting role as an extension of the existing representational space and enables multiple modes of communication. It acts as an extra layer, enabling us to involve more actors into the design studio education in a structured manner. As illustrated above, the theoretical component became an integral part of the design studio and the theory teachers (who are also architects/urban designers) became actively involved into the periodic critiques.

Furthermore, in this setting it is possible to share student works with the external experts and representatives of non-governmental organisations as well as students from different studios. Figure 2 represents this capacity.

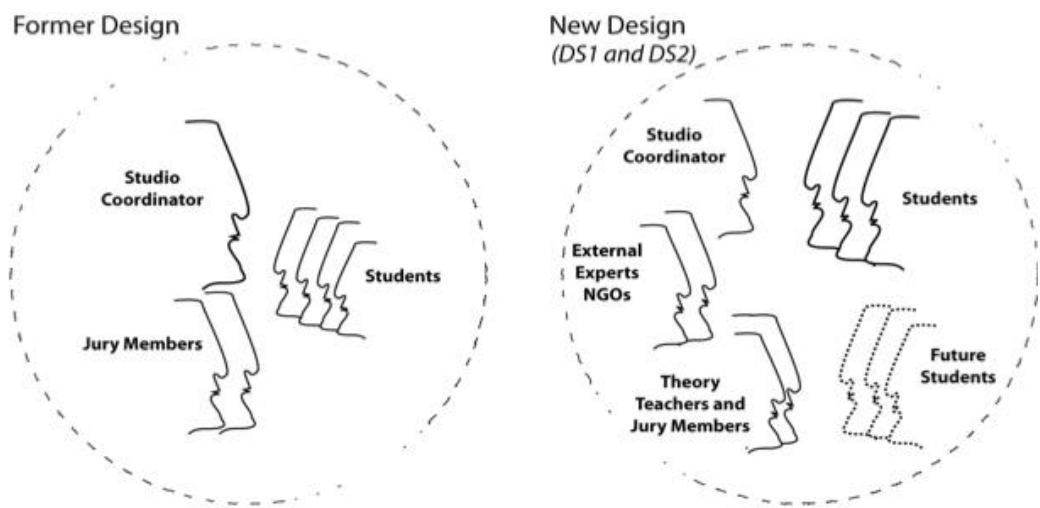

*Figure 2.* Actors involved in the conventional (on the left) and the experimental cases DS1 and DS2 (on the right).

In this sense, the two experimental urban design studios were not only configured to test a single platform, but with a holistic practice-based vision integrating social media and geographic information systems into the design studio in a novel manner. In both of the design studios, using a variety of open-source tools, technologies and in-house developed software, we enabled multiple methods of peer collaboration. These technologies were combined and tested in the past three years in real-life practices together with the participation of two urban planning focused non-governmental organisations operating in the Brussels Capital Region. With the contributions of BRAL (Brussels Environmental Council) and Green Belgium, they were thoroughly studied in terms of their user effectiveness, efficiency and satisfactoriness (Pak & Verbeke, 2012). The findings and knowledge learned from these practices were used while designing the setup illustrated below (Figure 3).

### The Technological Framework

The students were provided with various tools including a collective mapping interface (for analysis and sharing), a personal dashboard, a filtering and discovery tool (which works in coordination with other interfaces), image galleries, and a social discussion module which allows commenting and "liking" other students' work.

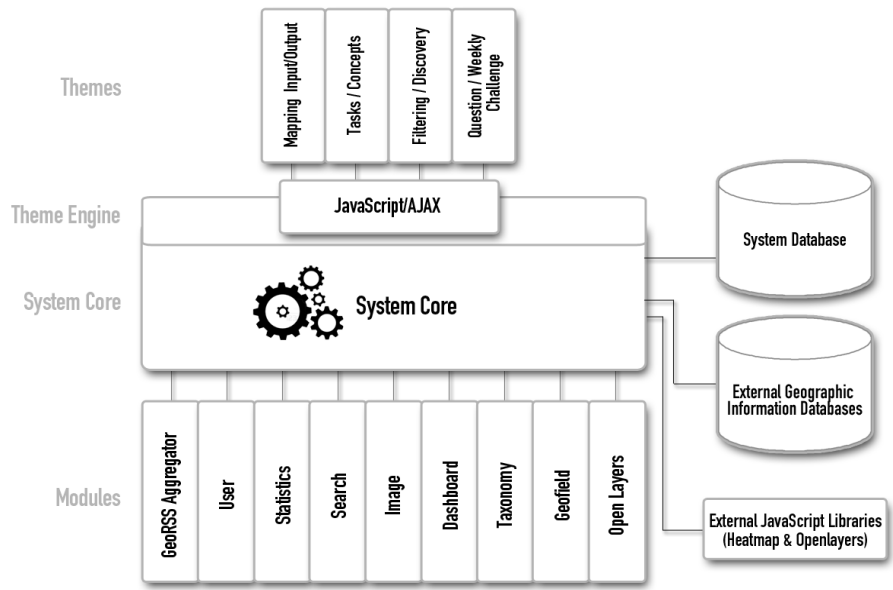

*Figure 3.*     The architecture of the social geographic web platform

In order to motivate the students, we assigned weekly tasks using the platform and asked them to collaboratively create maps of their urban analysis results and experiences, periodically upload their works and answer open-ended conceptual questions which were actually verbal diaries of their design processes.

### DS1: International Urban Design Studio 2012 (Project site: Luxembourg)

The DS1 studio was organised during the Spring Semester of 2012 with the participation of 34 international students from eleven different countries in Europe. The majority of the students had no previous knowledge of the project site - the plateau of Kirchberg in the city of Luxembourg.

The design site was originally designed to accommodate one of the three official seats of the European Union. Today, the relation between Kirchberg and the city centre is weak and the area lacks basic amenities such as cafes, restaurants and other sociable public spaces. For this reason, students were motivated to design solutions to improve the quality of life in the area, specifically focusing on eating-related urban spaces.

The students were divided in eight groups of four or five members. They were motivated to explore the inner city (especially the "eatscape") and use the results of their exploration as a source of inspiration.

The aim was to understand what makes those ordinary eat-related amenities "the places" in the city in a better way and how the locals themselves relate to those places through their own human situations, events, meanings, and experiences. The results of this exploration, represented in Figure 4, included design ranging from urban furniture to large scale urban re-thinking.

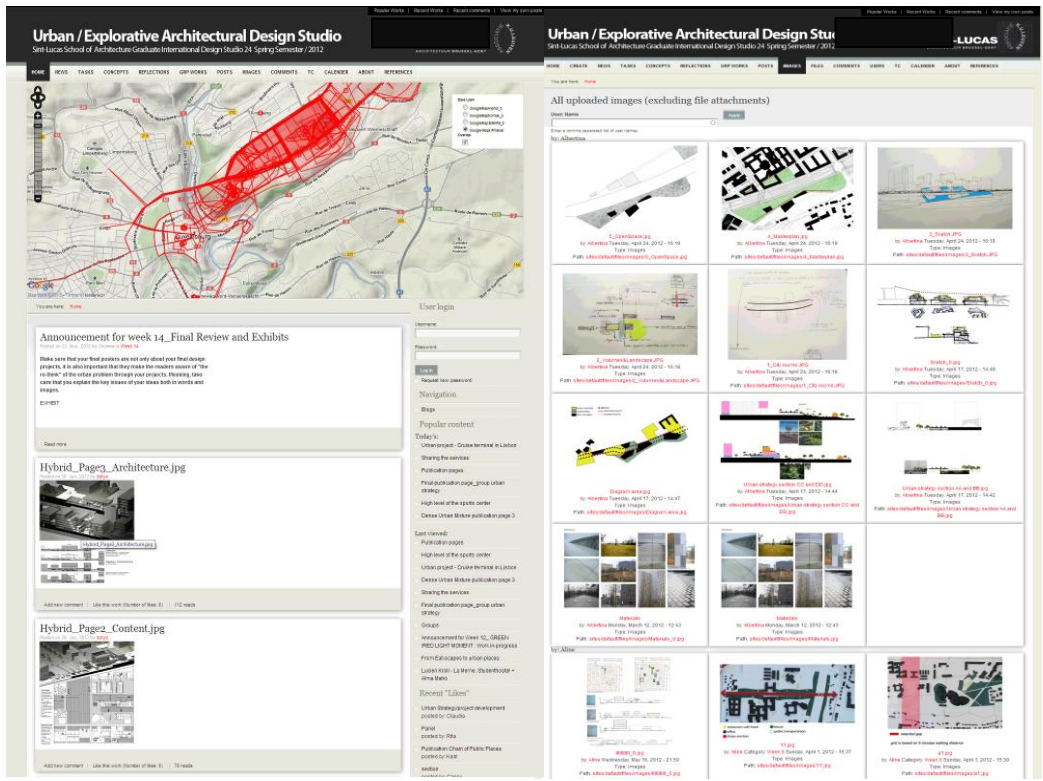

*Figure 4.*    The interface of the Social Geographic Web Platform used in DS1, Spring 2012.
http://www.emastudio2012.be

The students created a collective map of their own experiences using the web platform, overlaid with various types of external geographic information. These were maps from the GIS system of Luxembourg, *Google Maps*, *Open StreetMap* and *Bing* Maps. This feature served to the studio's aim of exploring the inner city of Luxembourg and using this as a source of inspiration. A dynamic map was allocated in the main page which superposes all of the explorations of the design students in a layered and interactive format (see Figure 5).

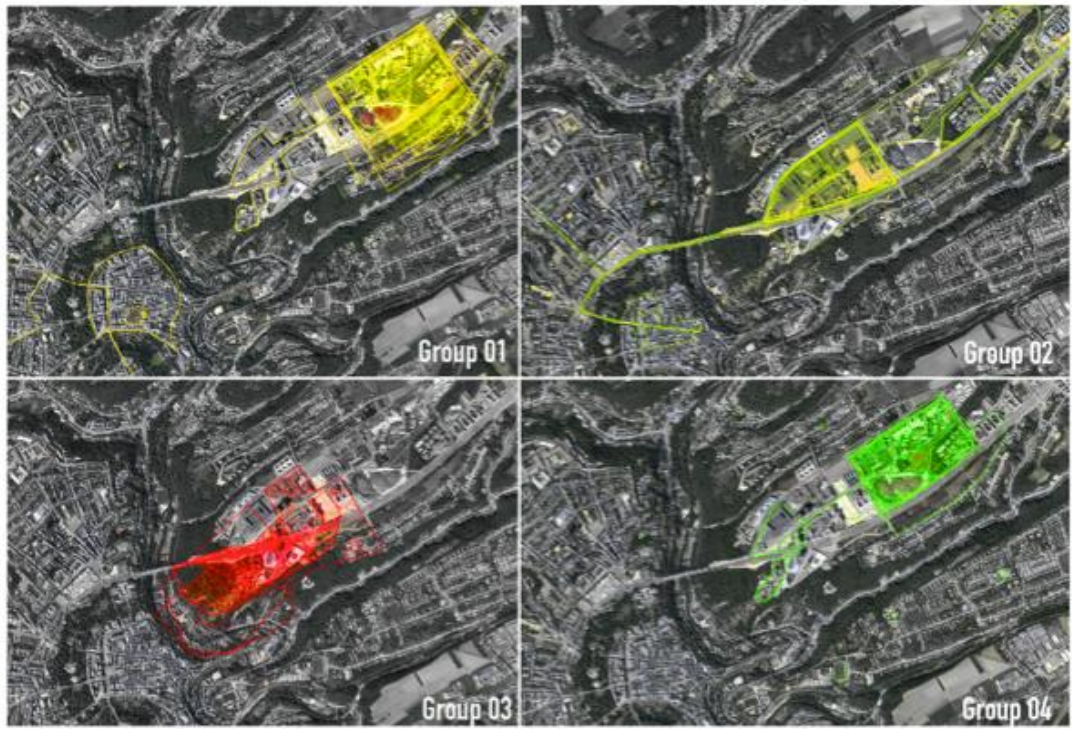

*Figure 5.* Four layers of the collective map created by the student groups indicating differences between focal points and serve as a background for the definition of the design problems.

### DS2: International Urban Design Studio 2013 (Project site: Brussels)

The second design studio took place during the Spring Semester of 2013. Twenty-seven international students from ten different countries in Europe participated in this course. The project site was located in the former Government Administrative Centre of Brussels. This area was chosen because it is a potential development area with significant socio-spatial problems which emerged as a part of the North-South railway connection passing through the historical centre.

The students worked in seven groups consisting of three or four members. They were asked to make a thorough collaborative urban analysis of the project site and shared their findings on the provided web platform in the form of responses to the challenges issued by the studio coordinators. Following the analysis phase, they explored alternative ways of reconfiguring the spatial landscape of the focus area in a designerly manner, using the collectively constructed body of knowledge.

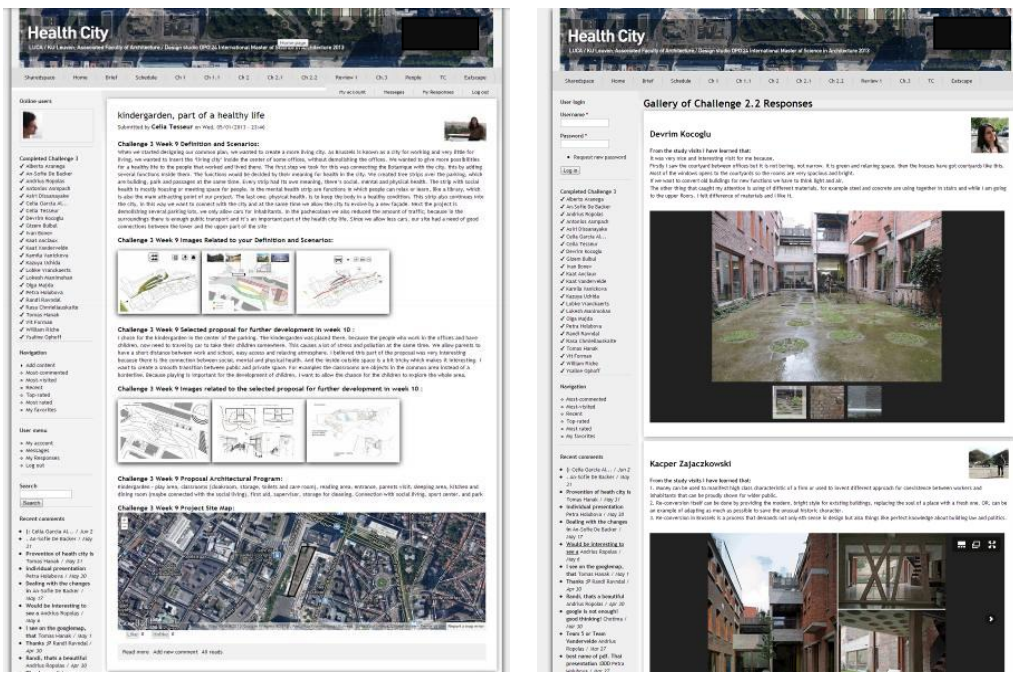

*Figure 6.* The interface of the Social Geographic Web Platform used in DS2, Spring 2013.
http://www.healthcity2013.be/

The web platform used in this case included various improvements in line with the feedback received in the first design studio experiment (DS1). First of all, the usability of the platform was improved by various readjustments and fixes. The students were given the opportunity to access the web platform using their personal smart/mobile phones. This allowed them to share their experiences on location, taking advantage of the location sensors installed in their smart phones and tablets. In addition, the student submissions in each challenge were displayed as image galleries which were used on demand during the critiques.

## Findings

This study is a result of an interactive inquiry process in which the first author was an active participant in the design studio as a teacher. In this sense, the knowledge generation in the two cases involved "a continuing reflection on practice under real-time conditions" (Argyris & Schön, 1974, p. 157). The experiences acquired from the first experimental studio (DS1) were reflected as improvements and readjustments onto the design of the web platform employed in the second studio (DS2).

As briefly described in the introduction, the intentions of our research were to develop a better understanding of the Design Studio 2.0. This included an investigation of:
- The nature of learning and participation that took place in the introduced Social-geographic Web 2.0 platform during the design studios DS1 and DS2;
- Perceived affordances of the platform and the students' views of its effectiveness, efficiency and satisfactoriness; and.
- Challenges and future development opportunities.

We sought answers to these questions during the course, relying on multiple data sources (Table 2). Among those were use logs collected through the proposed platform, student grades, an online questionnaire and a feedback meeting for each design studio. In addition to these, the first author took notes of the topics that arose during the process. The data was processed and sources were cross-compared while searching for the possible reasons for similar and conflicting observations. As a result we were able to distil several arguments which are revealed below.

Table 2
*Data sources used for the analysis of the experimental design studios.*

| Studio | Data Source | Explanation |
|---|---|---|
| **DS1** International Urban Design Studio 2012 | Online use logs Web-based Questionnaire Feedback Session Student Grades Notes | 15298 hits 15 Questions, 1 open-ended, 14 likert-scale (*n*= 22) 1 (*n*= 34) 1 (n= 34) |
| **DS2** International Urban Design Studio 2013 | Online use logs Web-based Questionnaire Feedback Session Student Grades Notes | 60356 hits 15 Questions, 1 open-ended, 14 likert-scale (n= 16) 1 (*n*=27) *1 (n=27)* |

### *The nature of learning and participation*

In the first experimental design studio (DS1), the students shared 611 design drawings and images relating to their design process, organised in 952 posts. In total, 15298 hits were logged onto the system, reflecting the intensity of use. During the second design studio (DS2), the students used the platform far more frequently recording 60356 hits - almost four times more than the first design studio. Accordingly, the number of drawings and images were 1092. In this studio (DS2), 513 posts were made, considerably fewer than for DS1. This was because the platform utilised in this case included significant usability improvements allowing sharing content in grouped galleries. This data is summarised in Table 3.

Table 3
*Logged student activities in DS1 and DS2*

| Studio | Students # | Hits # | Posts # | Drawings/Images Shared # |
|---|---|---|---|---|
| Design Studio DS1 | 34 | 15298 | 952 | 611 |
| Design Studio DS2 | 27 | 60356 | 513 | 1092 |

In order to search for the possible impacts of the platform on student learning, we have compared students' activities and grades. As introduced in the previous section, the functions and usability of the two platforms were different. We have noted a similar, almost logarithmic pattern in both cases (see Figure 7). The students' progress tended to increase together with their online activity up to a certain point. After this point, it stays flat and slowly starts to decrease at the end. In both cases (DS1 and DS2), this point is fairly close to the mean number of hits per person (the blue vertical line in Figure 7). All failed (grades less than 10; 9 with tolerance) students' online activity is less than twenty per cent of the total amount.

This observation suggests that the students' participation in the introduced web platform may be positively related to their progress in the Design Studio up to a certain limit. After this limit, this relationship starts to recede and produce negative effects.

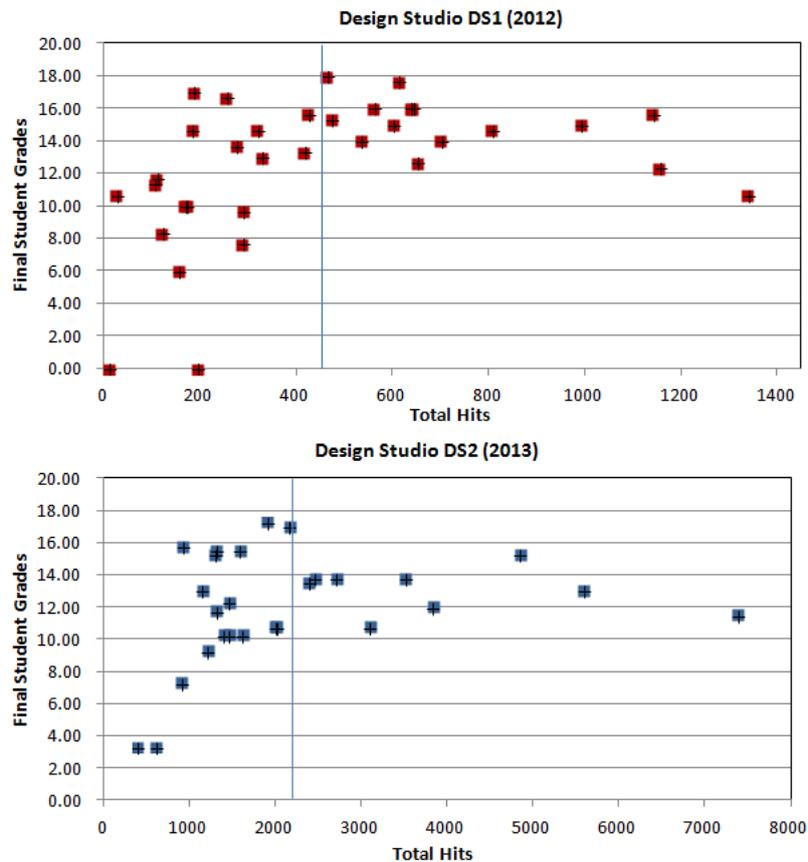

*Figure 7.* Comparison of online student participation with student grades as a learning indicator.
(The blue line indicates the mean number of hits per students)

Some of the plausible explanations of this effect can be:
- Over-participation may be shifting the focus towards the online platform itself rather than the design process and learning;
- Some of the students who are facing difficulties in the design process may be motivated to use the platform more often as a source of inspiration but fail to produce better results;
- Due to the individual differences between the students, the impact of online learning on student progress may be different; and,
- A combination of the above.

Another dimension of student participation was commenting on other students' contributions. By providing this opportunity, we aimed at motivating them to criticise and learn from each other through asynchronous conversations (as discussed in the introduction to this paper).

In the first design studio (DS1), 41 comments were made through the use of the platform. This result can barely be considered as successful since 11 students (*n*=34) did not post any comments on other students' works. In contrast, during the second studio (DS2) this number was 145. Increased usability of the new platform should have had a significant impact on this change.

A significant majority of the students (82%) made at least three comments (Figure 8).

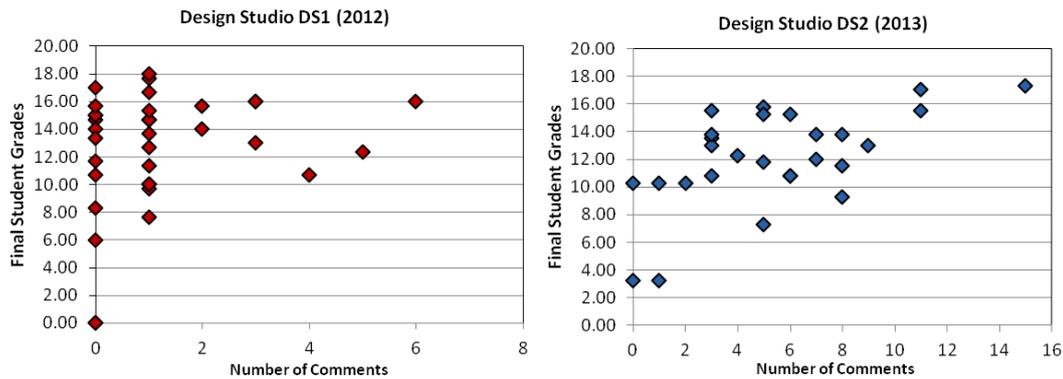

*Figure 8.* The comparison of the number of student comments with student grades

From the observations in Figure 8, we can conclude that making fewer comments did not necessarily cause the student to be less successful (although a weak relation is evident in the DS2 case). But it is making more comments than the average (DS1 *avg*=1.2, DS2 *avg*=5.37) is related to getting better grades. In DS1, all but one of the students who made one comment passed their classes (including a tolerable 9 of a possible 20). In DS2, all of the students who received higher grades than 10 (of a possible 20) made at least three comments.

### *Perceived Affordances and Benefits Experienced By the Students*

We received highly positive responses from the participating students during the feedback sessions and our online questionnaire, revealing some of the perceived potential and benefits (Table 4). Some of the students, for example, R1, noted the superiority of the introduced platform when compared with the existing e-learning system of the school based on the Blackboard (TOLEDO); which was not specifically tailored to be used in a design studio and is the general university platform for all faculties. Others addressed the benefits of being able to see other students' projects in a continuous manner, for example, R2 and R3. From these findings, we can conclude that the introduced platform was perceived to afford the (urban) analysis phase better than the final stages of project design.

Table 4
*Examples of positive reactions collected via our online questionnaire*

| Studio | Response |
|---|---|
| **Design Studio DS1** | **R1.** *It was a convenient way to handle the assignments and store your own files at the same time, much better than Toledo in any case. (Danail)*<br>**R2.** *Overall, the website was interesting and it was easy to create new posts, and it was nice to be able to see other people's works at any time, but at the same time, it was obvious that this kind of approach to teaching has to be done very carefully. (Ricardo)* |
| **Design Studio DS2** | **R3.** *Great job. Very valuable source in the course of a design studio: it's a very interactive and continuous learning (environment).(Celia)*<br>**R4.** *It was very handy to have a look at the projects of the other teams, at the beginning it was even very interesting to raise some discussions. At the end, when everybody were fully concentrated on their own projects, it became very formal just to fill the given task. Thought to conclude, the online platform was is very useful, also as the information source of the completed projects. (Rasa)* |

We asked the students several Likert-scale questions in order to collect more feedback on the perceived affordances. Three of these related to potentials of the web platform to contribute to learning from other students (Q1), the development of a better understanding of the project site (Q2) and learning from the external experts (Q3).

Figure 9 present the data from these questions from both DS1 and DS2. In both of the design studios, the majority of the students (DS1:78%, DS2:88%) strongly, mostly or somewhat agreed that they were able to learn from other students through the use of the web platform (Q1). Accordingly, the majority of the DS1 (75%) and the DS2 students (86%) responded positively to the question (Q2) relating to the potential of the web platform to facilitate a better understanding of the project site. Compared with DS1, the higher percentage of agreement in DS2 was a predictable result since sequential improvements were made to the web platform.

In contrast, the significant difference in the responses to the question (Q3) was unexpected. Over half of the students (59%) in DS1 disagreed that they were able to learn from the experts through the use of the web platform whereas this ratio in DS2 (25%) was considerably less. In order to explain the possible reasons for this observation, it is necessary to discuss the specific context in DS1. During the DS1 studio, an expert working for the city of Luxembourg made 14 comments on the students' work. The first author noted a shared sense of dissatisfaction among the students due to the contents of these comments. From this instance, we have learned that facilitating communication between students and experts does not necessarily support mutual learning. Learning is highly dependent on the profiles of the experts and their ability to express their opinions in a constructive manner. An additional finding on our feedback collection method was that the questionnaire was sensitive to student experiences and can also reflect negative perceptions.

**DS 1 (*n*=22)**

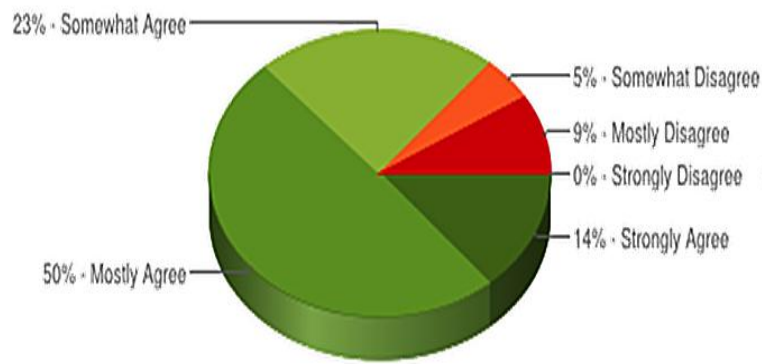

**DS 2 (*n*=16)**

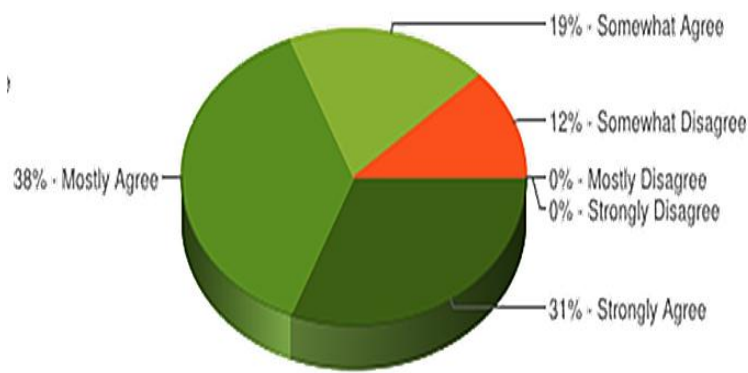

*Figure 9.* Student responses to the Likert scale questions in the online questionnaire

### Challenges and future development opportunities

During the feedback sessions, the students reported several types of challenges they faced and suggested various directions of improvement (Table 5). These can be grouped into four categories:
1. The complexity and organisation of design-related information;
2. The existing limitations of electronic visual displays;
3. Increased pressure on the students during the delivery of assignments; and,
4. Authorship and acknowledgement issues.

*Complexity and organisation of design-related information*: To start with, the complexity of the content produced by students was shown to be one of the most challenging issues. It was necessary to structure the large amount of text and visuals collected on the platform in an inspiring and easy-to-access form. In this context, the high amount of textual information on the web platform was reported as a barrier (R5). It was suggested that the platform had to be more visual, stressing the importance of visual thinking in the discipline (this discussion may relate to the dispositions of *Generation Y* (Strauss & Neil, 2000)). In line with this, two students (R7, R8) called for the simplification and better structuring of the interface by the rearrangement of the challenge topics and navigation as well as making some of the content private. This feedback can be useful for reducing the amount and complexity of the shared information.

*Limitations of electronic visual displays*: Secondly, due to the limited size and resolution of the electronic visual displays (monitors), it was not possible to organise online side-by-side poster exhibitions like we do in real life; which partially contributed to the difficulties in the management and retrieval of design information. There seems to be space for development in this direction.

Table 5

*Examples from the reactions collected through the open-ended question in our online questionnaire*

| Studio | Response |
| --- | --- |
| **Design Studio DS1** | **R5**. *The accent should be in any case on the visual information... If the website is more interactive and visual, it will surely be more interesting for the students and stimulating them to spend more time in it. (Danail)* |
| | **R6**. *I still enjoy the old method of just delivering things physically to the teacher, especially because like that we do not need to think about a few problems that may come from the website, for example: if our internet crashes; the 404 Error and so on. (Ricardo)* |
| **Design Studio DS2** | **R7**. *For the future development I would suggest to make the navigation easier by the topics and design progress levels. Also, it should be more compact, the long posts are good for evaluation and full understanding of ideas, but not comfortable for other users. (Rasa)* |
| | **R8**. *Lighter design. Easier system of finding materials, by type, team, person... Also maybe there could be public and private parts of website, so all in progress material would be not publicly available, just the few main challenges or workshops with really quality data. And for rest there could be even more in progress 'dirty' things, like sketches. (Andrius)* |

*Increased pressure on students during the delivery of assignments*. Another issue that was raised by the students was the additional stress during the delivery of the assignments due to possible technical issues and the limited Internet connection speed. In this case, handing drawings physically to the teachers was perceived as a comparatively convenient method free from technical problems or errors (R6). It is apparent that the accumulation of these issues can have a significant negative impact on participation and performance.

*Authorship and acknowledgement issues*: Besides the issues identified through our questionnaire, *Authorship* emerged as an important problem during the feedback session of the design studio DS1. Student (R9) expressed that she was *"highly uncomfortable with another student (R10) who excessively used her ideas and integrated them into her own design."* After a quick review of the shared drawings on the web platform and their dates, it became clear that the main concept and a few ideas, had been "borrowed" by R10. These included *"linking the two sides of the valley of Kirchberg"* and *"a building in the valley with a curved roof flowing with the slope."*

During the reviews, these similarities were perceived as normal by the teachers since it is quite common for two students to come up with similar ideas. Being inspired by another student is acceptable; and the Design Studio 2.0 setup motivates students towards this direction. In this case, the problem was the lack of acknowledgement. From this experience, we learned that while encouraging design students to learn from each other, we also needed to inform them about possible authorship issues that may arise and remind them to give credit where it was due. At the beginning of the DS2, while introducing the web platform, we gave this advice to the students and no similar problems were observed.

## Conclusion

In this paper, we have discussed the opportunities provided by the combination of Web 2.0, social media and geographic technologies for learning and for new forms of participation in an urban design studio. We based our arguments on our recent findings from two experimental urban Design Studio 2.0 setups and former studies focusing on this topic.

We revealed our observations relating to the nature of learning and participation that took place in these studios as well as the perceived affordances and challenges of using Web 2.0 social geographic platforms in the design studio, drawing on the learning experiences of the students. We received highly positive responses during the feedback sessions and the online questionnaires. The students found the platform convenient and addressed it as a valuable resource for learning.

Reflecting on our own experiences, the web platform enabled us to extend the learning that took place in the design studio beyond the studio hours, to represent the design information in novel ways and allocate multiple communication forms.

Through the introduced setup and the web platform, we were able to augment urban design learning; remediate and extend the reflective conversation in the design studio. Using collaborative mapping functionality, the students collectively constructed a shared memory of urban spaces which reportedly helped them to develop a better understanding of their project site. They were able to learn from their peers as well as the external experts.

Moreover, it was possible to combine conventional and online learning activities. By this way, the focus of the design studio was oriented more towards the students and the learning processes. The students frequently commented on each other's works and constructed a common understanding. From a critical point of view, it is important to note that almost all of the student comments on each other's work were positive. The students refrained themselves from making negative comments on a public web platform.

In addition to the above, through the analysis of the usage logs, we found that the students' participation in the introduced web platform may be positively related to their progress up to a certain point. The relation is clear, but the direction of causality is still a question mark.

In both of the design studios (DS1 and DS2), the students with the highest online activity barely passed the course (Figure 7). A plausible explanation of this observation can be over-participation. Due to excessive use of the platform, the focus of the students may be shifting towards the medium itself rather than the design process. It is possible that the students who have difficulties in coming up with novel creations may be seeking more information or support. More research on the students' individual learning and design styles is needed to clarify this issue.

For the future studies, we need to consider the possible negative effects of reported over-participation. During this process, the individual differences between the students should be addressed carefully, since the impact of online learning on student progress can vary based on their backgrounds. In parallel, it is necessary to create novel technological solutions to reduce the complexity of the design information and find alternative integrated strategies to promote online and offline constructive reflection-in-action.

**Acknowledgements**

This research is partially based on a three year postdoctoral research project supported by the Brussels Capital Regional Government, Institute for the Encouragement of Scientific Research and Innovation (INNOVIRIS) given to Dr Burak Pak, promoted by Dr Johan Verbeke. The authors would like to thank the students for their contributions and completing the questionnaires.